\documentclass[12pt]{article}
\usepackage{amsmath}

\input{epsf}
\def\lra{\leftrightarrow}

\newcommand\as{\alpha_{\mathrm{S}}}
\newcommand\gs{g_{\mathrm{S}}}
\newcommand\eij{e^{-i \lambda_{ij} \pi}}
\newcommand\eiq{e^{-i \lambda_{iq} \pi}}
\newcommand\ejq{e^{-i \lambda_{jq} \pi}}

\newcommand\f[2]{\frac{#1}{#2}}
\def\ep{\epsilon}
\newcommand\vep{\varepsilon}

\def\ee{$e^+e^-$}
\def\beeq{\begin{eqnarray}}
\def\eeeq{\end{eqnarray}}
\def\cm{{\cal M}}
\def\pb{{\bar p}}
\def\bom#1{{\mbox{\boldmath $#1$}}}
\def\to{\rightarrow}

\newcommand{\la}{\langle}
\newcommand{\ra}{\rangle}

\def\nn{\nonumber}

\def\ID{1 \kern -.45 em 1}

\def\ket#1{|{#1}\ra}
\def\bra#1{\la{#1}|}

\begin{document}

\begin{titlepage}
\renewcommand{\thefootnote}{\fnsymbol{footnote}}
\begin{flushright}
     CERN--TH/2000--184 \\ hep-ph/0007142
     \end{flushright}
\par \vspace{10mm}
\begin{center}
{\Large \bf
The soft-gluon current at one-loop order~\footnote{This work was supported 
in part 
by the EU Fourth Framework Programme ``Training and Mobility of Researchers'', 
Network ``Quantum Chromodynamics and the Deep Structure of
Elementary Particles'', contract FMRX--CT98--0194 (DG 12 -- MIHT).}}
\end{center}
\par \vspace{2mm}
\begin{center}
{\bf Stefano Catani}~\footnote{On leave of absence from INFN,
Sezione di Firenze, Florence, Italy.}\\

\vspace{5mm}

{Theory Division, CERN, CH-1211 Geneva 23, Switzerland} \\

\vspace{5mm}

{and}

\vspace{3mm}

{\bf Massimiliano Grazzini}~\footnote{Work supported by the Swiss
  National Foundation.}\\

\vspace{5mm}
Institute for Theoretical Physics, ETH-H\"onggerberg, CH-8093 Zurich, 
Switzerland

\vspace{5mm}

\end{center}

\par \vspace{2mm}
\begin{center} {\large \bf Abstract} \end{center}
\begin{quote}
\pretolerance 10000

We study the soft limit of one-loop QCD amplitudes and we derive the 
process-independent 
factorization formula that controls the 
singular behaviour
in this limit. This
is obtained from the customary 
eikonal factorization formula valid at tree (classical) level 
by introducing a generalized soft-gluon current that embodies the quantum
corrections.
We compute the explicit expression of the soft-gluon current at one-loop order.
It contains purely non-abelian correlations between the colour charges of each
pair of hard-momentum partons in the matrix element. This leads to
colour correlations between (two and) three hard partons in the matrix element
squared. Exploiting colour conservation, we recover QED-like factorization for
the square of the matrix elements with two and three hard partons.

\end{quote}

\vspace*{\fill}
\begin{flushleft}
     CERN--TH/2000--184 \\June 2000 

\end{flushleft}
\end{titlepage}

\renewcommand{\thefootnote}{\fnsymbol{footnote}}
\section{Introduction}
\label{sec:intro}

Higher-order computations in
perturbative
QCD
can be performed by using
three main tools: exact calculations at a fixed order in the QCD coupling 
$\as=\gs^2/4\pi$, analytic resummed calculations and parton shower event 
generators. At present, the accuracy of these tools is respectively limited 
to the next-to-leading order (NLO), to next-to-leading logarithmic (NLL)
accuracy and to the dominant soft and collinear enhanced contributions.
An extensive and updated list of references can be found in
Refs.~[\ref{Catani:2000jh}, \ref{Catani:2000zg}].

Apart from
important conceptual and technical differences, these tools are based on a
common ingredient: the universal factorization properties of QCD amplitudes in
the infrared (soft and collinear) region. The lowest-order version of the 
soft and collinear factorization formulae [\ref{AP}, \ref{BCM}] 
was indeed exploited [\ref{book}] to develop these tools to their
present theoretical accuracy. Higher-order versions of the factorization 
formulae are required to progress towards next-to-next-to-leading order
(NNLO) calculations, resummation of next-to-next-to-leading logarithmic (NNLL)
terms and inclusion of subdominant contributions in parton showers.

In recent years several groups have contributed to extending infrared 
factorization to higher perturbative orders [\ref{glover}--\ref{sing2loop}].

The general factorization properties of tree-level and loop amplitudes
in the limit where two or more partons become collinear were studied in 
Refs.~[\ref{glover}--\ref{1loopepskos}]. At tree level,
the singular factors for the collinear splitting of
one parton into
three 
were explicitly computed in Refs.~[\ref{glover}, \ref{Catani:2000ss},
\ref{Catani:1999nv}].
The splitting function
corresponding to the clustering
of four collinear gluons is also 
known [\ref{DelDuca:2000ha}]. The one-loop kernels for 
the collinear splitting of
one parton into
two were obtained in 
Refs.~[\ref{Bern:1998sc}--\ref{Bern:1999ry}].

The mixed soft-collinear limit [\ref{glover}, \ref{Catani:2000ss}] 
can be studied by exploiting the {\em coherence} properties of QCD 
radiation [\ref{BCM}]. Using QCD coherence, the singular behaviour in the
soft-collinear limit can be treated [\ref{Catani:2000ss}]
by combining the singular factors that separately control the collinear and 
soft limits.

The soft limit is physically more involved than the collinear limit.
Long-wavelength (soft) gluons can spread the colour flow over large 
distances, thus leading to (non-local) colour correlations. Kinematics and 
colour factors turn out to be deeply entangled in the soft-factorization
formulae.

The tree-level factorization formulae for the emission of two soft gluons
were independently derived in Refs.~[\ref{bgdsoft}] and [\ref{Catani:2000ss}].
The soft (and collinear) singular behaviour of two-loop amplitudes was
studied in Ref.~[\ref{sing2loop}].

In this paper we consider the limit in which a soft gluon is radiated
from one-loop amplitudes. The limit was first investigated in 
Refs.~[\ref{Bern:1995ix}, \ref{Bern:1998sc}, \ref{Bern:1999ry}].
The formalism used by this group is based on the decomposition 
of the one-loop matrix elements in colour subamplitudes [\ref{1loopdec}]. 
The colour-subamplitude decomposition depends on the type of external
partons, and these authors derived the explicit expressions of the one-loop
soft-gluon contribution to colour subamplitudes with $m$ external gluons 
[\ref{Bern:1998sc}] and with $m$ external gluons plus a $q{\bar q}$ pair
[\ref{Bern:1999ry}].

In the present
paper
the soft limit is studied by means of a completely
independent and general method. We apply the eikonal approximation and 
soft-gluon insertion rules to perform infrared factorization directly
in colour space. Within
this
formalism, the soft limit of 
tree-level amplitudes is described by a factorization formula written in terms
of a soft-gluon (or eikonal) current (see Ref.~[\ref{BCM}] and 
Sect.~\ref{sec:softlo}) that describes colour radiation in the classical
approximation. We show that the factorization formula can be extended
to loop amplitudes by introducing a generalized soft-gluon current that 
embodies (non-abelian) quantum corrections. As at tree level, 
the soft-gluon current
only depends on the colour charges and momenta of the external partons
in the loop amplitude. We compute the explicit expression of
the soft-gluon current at one-loop order.

Our results for the soft limit of one-loop amplitudes agree with those 
in Refs.~[\ref{Bern:1998sc}, \ref{Bern:1999ry}] 
for the particular cases considered therein. 
The one-loop gluon current that we obtain is derived and presented in a general
and process-independent way; it can be applied to any one-loop amplitude.
In particular, we can easily show that colour and kinematic factors 
can be completely disentangled in the computation of
the soft limit of the {\em square} of one-loop matrix elements
with two or three external QCD partons. This simplified factorization structure
is particularly useful for the NNLO calculation of 2-jet and 3-jet cross
sections in $e^+e^-$ annihilation.  

Infrared-factorization properties at the lowest perturbative order have been known
for a long time [\ref{BCM}]. Combining the results obtained here with those 
in Refs.~[\ref{glover}--\ref{sing2loop}], the general structure of the
infrared singularities at the next perturbative order
is also completely and explicitly known. It can be used to improve the 
accuracy of perturbative QCD calculations.

The paper is organized as follows. In Sect.~\ref{sec:softlo} we recall the known
results for the soft limit of tree-level amplitudes. In Sect.~\ref{sec:1loopfac}
we present the process-independent 
factorization formula at higher perturbative orders and we discuss in detail 
its features at one-loop order. In
Sect.~\ref{sec:1loopcur} we prove the factorization formula at one-loop order
and we derive the explicit expression of the one-loop soft current. In
Sect.~\ref{sec:examples} we apply our results to the squared amplitudes of
processes with two or three hard partons.

\section{Soft-gluon factorization at tree level}
\label{sec:softlo}

We consider a generic scattering process that involves $m$ external QCD partons
({\em massless} quarks and gluons) with momenta $p_1, \dots, p_m$ 
and an arbitrary number and type of particles with no colour 
(photons, leptons, vector bosons, ...).
Note that, by definition, we always consider incoming and outgoing parton 
momenta in the physical region, i.e. any $p_i$ is massless, with 
positive-definite energy (in particular, $p_i\cdot p_j > 0$).
The corresponding matrix element is denoted by $\cm(p_1,\dots,p_m)$
and the dependence on the momenta and quantum numbers of 
non-QCD particles is always understood.

The matrix element has the following loop expansion:
\begin{equation}
\label{loopexp}
\cm(p_1,\dots,p_m) = \cm^{(0)}(p_1,\dots,p_m) + \cm^{(1)}(p_1,\dots,p_m)
+ \dots \;\;,
\end{equation}
where $\cm^{(0)}$ denotes the tree-level contribution, $\cm^{(1)}$ denotes the
one-loop contribution, and the dots stand for higher-loop corrections.
Note that we always consider unrenormalized matrix elements. Thus 
Eq.~(\ref{loopexp}) is a power series expansion in the bare QCD coupling
$\gs$ and, in particular, $\cm^{(1)}$ is the unrenormalized one-loop
amplitude.

We simultaneously regularize ultraviolet and infrared singularities by using
dimensional regularization. Apart from introducing the 
dimensional-regularization scale $\mu$ through the replacement 
$\gs \to \gs \,\mu^\ep$,
the key ingredient of dimensional regularization is
the analytic continuation of loop momenta to $d=4-2\ep$ space-time dimensions.
Having done this, we are left with some
freedom regarding the dimensionality of the momenta of the external
particles as well as the number of polarizations of both external and
internal particles. This leads to different regularization schemes 
[\ref{tHV}--\ref{4dhs}]
within the dimensional-regularization prescription. 
The regularization-scheme dependence of one-loop amplitudes was studied
in detail in Refs.~[\ref{KST2to2}, \ref{uni}]. 
All the results on the 
soft behaviour presented in this paper do not explicitly depend
on the dimensional-regularization scheme 
(see Sect.~\ref{sec:1loopcur} 
for a brief discussion of different regularization schemes): 
the scheme
dependence is implicitly embodied in the expressions of the tree-level
and one-loop matrix elements $\cm^{(0)}$ and $\cm^{(1)}$.

The emission of a soft gluon does not affect the momenta and spins
of the radiating hard partons. However, it does affect their colour because 
the gluon always carries away some colour charge, no matter how soft it is. 
Unlike the case of soft-photon emission in QED, soft-gluon emission thus
does not factorize exactly and leads to colour correlations.

To take into account the colour structure without referring to any particular
choice of basis colour vectors
(such as, for instance, the decomposition in colour subamplitudes 
[\ref{Bern:1999ry}]), we use a general notation (see e.g. 
Ref.~[\ref{CSdipole}]).
The dependence of the matrix element on the colour indices $c_1,\dots,c_m$
of the QCD partons is written as
\begin{equation}
\label{cmmdef}
\cm_{c_1,\dots,c_m}(p_1,\dots,p_m) \equiv \la c_1,\dots,c_m \,|\,
\cm(p_1,\dots,p_m) \ra \;.
\end{equation}
Thus $\{\,\ket{c_1,\dots,c_m} \,\}$ is an abstract basis in colour space and 
the ket $\ket{\cm(p_1,\dots,p_m)}$ is a vector in this space.
According to this notation, the matrix element squared $|\cm|^2$ (summed
over the colours and spins of the partons) can be written as
\begin{equation}
|\cm(p_1,\dots,p_m)|^2 = 
\la \, \cm(p_1,\dots,p_m) \,
 | \, \cm(p_1,\dots,p_m) \, \ra \;.
\end{equation}

To describe the colour correlations produced by soft-gluon emission,
we associate a colour charge ${\bom T}_i$
with the emission of a gluon from each parton $i$. If the emitted gluon
has colour index $a$ ($a= 1, ...,$ $N_c^2-1$), the colour-charge operator is:
\begin{equation}
{\bom T}_i \equiv \bra{a} \;T_i^a 
\end{equation}
and its action onto the colour space is defined by
\begin{equation}
\bra{a,c_1,\dots, c_i, \dots, c_m} {\bom T}_i
\ket{b_1, \dots, b_i, \dots, b_m} = \delta_{c_1 b_1} ...
T_{c_i b_i}^a ...\delta_{c_m b_m} \;\;,
\end{equation}
where $T_{c b}^a \equiv i f_{cab}$ (colour-charge matrix
in the adjoint representation)  if the emitting parton $i$
is a gluon and $T_{\alpha \beta}^a \equiv t^a_{\alpha \beta}$
(colour-charge matrix in the fundamental representation with
$\alpha, \beta =1,\dots,N_c$)
if the emitting particle $i$ is a {\em final-state} quark or an
{\em initial-state} antiquark 
($T_{\alpha \beta}^a \equiv {\bar t}^a_{\alpha \beta}
= - t^a_{\beta \alpha }$, in the case of a final-state
antiquark or an initial-state quark).

The colour-charge algebra is\footnote{More details on the colour algebra 
and useful colour-matrix relations can be found in Appendix~A of 
Ref.~[\ref{CSdipole}].}:
\begin{equation}
T_i^a \, T_j^a \equiv
{\bom T}_i \cdot {\bom T}_j ={\bom T}_j \cdot {\bom T}_i \;\;\;\;{\rm if}
\;\;i \neq j; \;\;\;\;\;\;{\bom T}_i^2= C_i,
\end{equation}
where $C_i$ is the Casimir operator, i.e.
$C_i=C_A=N_c$ if $i$ is a gluon and $C_i=C_F=(N_c^2-1)/2N_c$ if $i$ is a quark
or antiquark. 

Note that, by definition, each vector $\ket{\cm(p_1,\dots,p_m)}$
is a colour-singlet state. Therefore colour conservation is simply
\begin{equation} 
\label{cocon}
\sum_{i=1}^m {\bom T}_i \; \ket{\cm(p_1,\dots,p_m)} = 0 \;.
\end{equation}

We can now recall the behaviour of the tree-level matrix element
$\cm^{(0)}(q,p_1,\dots,p_m)$ in the limit where the 
momentum $q$ of the gluon becomes soft. Denoting by
$a$ and $\vep^\mu(q)$ the colour and the polarization vector of the soft gluon,
the matrix element fulfils the following factorization formula~[\ref{BCM}]
\begin{equation}
\label{eikfac}
\la a \,|\, \cm^{(0)}(q,p_1,\dots,p_m) \rangle 
\simeq \gs \, \mu^\ep
\vep^\mu(q) \, J_{\mu}^{a \,(0)}(q)
\; |\, \cm^{(0)}(p_1,\dots,p_m)\rangle \;,
\end{equation}
where $|\, \cm^{(0)}(p_1,\dots,p_m)\rangle$
is obtained from the original matrix element by simply removing the soft gluon 
$q$. The factor ${\bom J}^{(0)}_\mu(q)$ is the tree-level soft-gluon current
\begin{equation}
\label{eikcur}
{\bom J}^{\mu \,(0)}(q)=\sum_{i=1}^{m} {\bom T}_i\,\f{p_{i}^{\mu}}{p_i\cdot q} \;,
\end{equation}
which depends on the momenta and colour charges of the
hard partons in the matrix element on the right-hand side of 
Eq.~(\ref{eikfac}). The symbol `\,$\simeq$\,' means that on the right-hand side
we have neglected contributions that are less singular than $1/q$ in the soft
limit $q \to 0$. Note that Eq.~(\ref{eikfac}) is valid in any number
$d=4-2\ep$ of space-time dimensions, and the sole dependence on $d$ is in the 
overall factor $\mu^\ep$.

The factorization formula (\ref{eikfac}) can be derived in a simple way
by working in a {\em physical gauge} and using the following
{\em soft-gluon insertion rules}. The coupling of the gluon to any 
{\em internal} (i.e. highly off-shell) parton in the amplitude 
$\cm^{(0)}(q,p_1,\dots,p_m)$ is not singular in the soft limit;
it can thus be neglected. The soft-gluon coupling to any {\em external}
or, in general, {\em nearly on-shell} parton with colour charge $\bom T$
and momentum $p$ can be factorized by using the {\em eikonal approximation},
that is by extracting the contribution 
$g_S \mu^\ep 2 p^\mu {\bom T}$ for the vertex and the contribution 
$1/(p+q)^2 \simeq 1/(p^2 + 2p\cdot q)$ for the propagator. Note that the 
eikonal vertex only depends on the momentum and colour charge of the radiating
parton: it does not depend on either the soft momentum or the spin of the
parton. Using the eikonal propagator simply amounts to neglecting the terms that are
quadratic in the soft momentum.  

An important property of the soft-gluon current is current conservation. 
Multiplying Eq.~(\ref{eikcur}) by $q^\mu$, we obtain
\begin{equation}
\label{eikcons}
q^\mu {\bom J}^{(0)}_\mu(q) = \sum_{i=1}^{m} {\bom T}_i \;,
\end{equation}
and thus
\begin{equation}
\label{curcon}
q^\mu {\bom J}^{(0)}_\mu(q) |\,\cm^{(0)}(p_1,\dots,p_m)\rangle
= \sum_{i=1}^{m} {\bom T}_i \;|\,\cm^{(0)}(p_1,\dots,p_m)\rangle
= 0 \;\;,
\end{equation}
where the last equality follows from colour conservation 
as in Eq.~(\ref{cocon}). Although the factorization formula (\ref{eikfac})
is most easily derived by working in a physical gauge,
the conservation of the soft-gluon current implies that Eq.~(\ref{eikfac})
is actually gauge invariant. Any gauge transformation is equivalent
to an addition of a longitudinal component to the polarization vector of the soft gluon
through the replacement $\vep^\mu(q) \to \vep^\mu(q) + \lambda q^\mu$.
Nonetheless the factorization formula (\ref{eikfac}) is invariant under this
replacement, because of Eq.~(\ref{curcon}).

Squaring Eq.~(\ref{eikfac}) and summing over the gluon polarizations leads to
the well-known soft-gluon factorization formula at ${\cal O}(\gs^2)$
for the squared amplitude [\ref{BCM}]:
\beeq
\label{ccfact}
| \cm^{(0)}(q,p_1,\dots,p_m) |^2 \simeq
- \gs^2 \,\mu^{2\ep} \,2 \,\sum_{i,j=1}^m\, {\cal S}_{ij}(q) 
\;| \cm^{(0)}_{(i,j)}(p_1,\dots,p_m) |^2 \;\;,
\eeeq 
where the eikonal function ${\cal S}_{ij}(q)$ 
can be written in terms of two-particle sub-energies 
$s_{ij}=(p_i + p_j)^2$ as follows
\begin{equation}
\label{eikfun}
{\cal S}_{ij}(q) = \f{p_i \cdot p_j}{2 (p_i \cdot q)\, (p_j\cdot q)}
= \f{ s_{ij}}{s_{iq} \,s_{jq}} \;\;.
\end{equation}
The colour correlations produced at tree level by the emission of a
soft gluon are taken into
account by the square of the colour-correlated tree-amplitude 
$| \cm^{(0)}_{(i,j)} |^2$ on the right-hand side. This is given by
\beeq
\label{colam}
| \cm^{(0)}_{(i,j)}(p_1,\dots,p_m) |^2 \!\!\!&\equiv&\!\!\!
\la \,\cm^{(0)}(p_1,\dots,p_m) \,|
\,{\bom T}_i \cdot {\bom T}_j 
\,|\,\cm^{(0)}(p_1,\dots,p_m)\rangle
\\
\!&=&\!\!\! \left[ 
\cm^{(0)}_{c_1.. b_i ... b_j ... c_m}(p_1,\dots,p_m) \right]^*
\; T_{b_id_i}^a \, T_{b_jd_j}^a
\; \cm^{(0)}_{c_1.. d_i ... d_j ... c_m}(p_1,\dots,p_m) \;.
\nonumber
\eeeq

\section{Soft-gluon factorization at one loop}
\label{sec:1loopfac}

The factorized structure of the soft limit of the 
QCD amplitudes at tree level can be generalized to higher loops.
The soft limit of the one-loop amplitudes is studied in detail in 
Sect.~\ref{sec:1loopcur}. In this section we anticipate and discuss 
the final results.

Our analysis is consistent with the following factorization formula
\begin{equation}
\label{genfac}
\la a \,|\, \cm(q,p_1,\dots,p_m) \rangle 
\simeq \vep^\mu(q) \, J_{\mu}^{a}(q,\ep)\; |\, \cm(p_1,\dots,p_m)\rangle 
\;\left[ 1 + {\cal O}(\gs^4) \right]  \;,
\end{equation}
where the symbol `\,$\simeq$\,' 
has the same meaning as in Eq.~(\ref{eikfac}).
The matrix element on the right-hand side is the all-loop
amplitude in Eq.~(\ref{loopexp}) and the singular dependence on $q$
is embodied in the (unrenormalized) soft-gluon current $J_{\mu}^{a}(q,\ep)$, 
which can be expanded in loop contributions, i.e. in powers of $\gs^2$:
\begin{equation}
\label{gencur}
J_{\mu}^{a}(q,\ep) = \gs \, \mu^\ep \left[ J_{\mu}^{a \,(0)}(q) + 
\gs^2 \, \mu^{2\ep} J_{\mu}^{a \,(1)}(q,\ep) + \dots \right] \;.
\end{equation}
The term $J_{\mu}^{a \,(0)}(q)$ is the tree-level current in Eq.~(\ref{eikcur}),
the term $J_{\mu}^{a \,(1)}(q,\ep)$ is its one-loop correction, and so forth.

The soft current contains the entire singular dependence in the soft limit and
no further approximation is performed in Eq.~(\ref{genfac}).
By this we mean that $J_{\mu}^{a}(q,\ep)$ behaves as $1/q$ 
(such as at tree level)
modulo any possible enhancement proportional to
powers of $\ln q$ coming from the loop contributions. In particular, the
$\ep$-dependence of the current can be evaluated exactly without performing any
$\ep$-expansion and, thus, by keeping all the powers of $\ln q$ coming from 
higher-loop contributions of the type $(q)^{\ep} = 1 + \ep \ln q + \cdots$.

The discussion in Sect.~\ref{sec:1loopcur} suggests that the factorization
formula (\ref{genfac}) is valid to any loop order. Nonetheless,
we have included the term ${\cal O}(\gs^4)$ on the right-hand side of 
Eq.~(\ref{genfac}) to indicate that
our explicit proof and calculation do not extend beyond the one-loop order.

Expanding both sides of Eq.~(\ref{genfac}) to one-loop accuracy and using
Eq.~(\ref{eikfac}), we can obtain
the factorization formula for the soft limit of the one-loop amplitudes:
\beeq
\label{1loopff}
\la a |\, \cm^{(1)}(q,p_1,\dots,p_m) \rangle \simeq 
g_S \, \mu^\ep \,\vep^\mu(q) 
\!\!\!&&\!\!\! \!\!\!\left[ \;J_{\mu}^{a \,(0)}(q) 
\;|\,\cm^{(1)}(p_1,\dots,p_n)\rangle
\right. \nn \\
&&+ \left. \gs^2 \, \mu^{2\ep} \,
J_{\mu}^{a \,(1)}(q, \ep) \;|\,{\cal M}^{(0)}(p_1,\dots,p_n)\rangle \;
\right] \;.
\eeeq
The explicit expression for the one-loop current is
\begin{align}
J_{a}^{\mu \,(1)}(q, \ep) &= -\f{1}{16\pi^2}\; \f{1}{\ep^2} \;\f{\Gamma^3(1-\ep)
\,\Gamma^2(1+\ep)}{\Gamma(1-2\ep)} \nonumber \\
\label{1loopcur}
&\,\cdot \;\; i\, f_{abc}\sum_{i\neq j}T_i^b\;T^c_j
\left(\f{p_{i}^{\mu}}{p_i\cdot q}-\f{p_{j}^{\mu}}{p_j\cdot q}\right)
\left[\f{4\pi\, p_i\cdot p_j \,\eij}{2 (p_i\cdot q)\, (p_j\cdot q) \, 
\eiq \,\ejq} \right]^\ep\, .
\end{align}
We remind the reader that in our notation all the incoming and outgoing momenta
are in the physical region (any $p_i$ has positive-definite energy and
$p_i \cdot p_j > 0$). Thus the complex factors $e^{-i \pi \lambda_{AB}}$ 
($\lambda_{AB}=+1$ if $A$ and $B$ are both incoming or outgoing, and
$\lambda_{AB}=0$ otherwise) in
Eq.~(\ref{1loopcur})
are the unitarity phases related to the
analytic continuation from unphysical to physical momenta.

The result in Eq.~(\ref{1loopcur}) explicitly
shows that the soft limit of the one-loop amplitudes is process-independent,
meaning that
it does not depend on the momentum and colour flows of 
the internal partons (including the parton circulating in the loop) in the
matrix element. This simple structure has several
interesting features that we comment below.

The one-loop soft current is proportional to the structure constants $f_{abc}$
of the gauge group, and thus it is purely {\em non-abelian}. 
This is in agreement with the absence of higher-loop corrections 
to the soft current in massless QED [\ref{YFS}].

Since Eq.~(\ref{1loopcur}) is proportional to the factor
$(p_{i}^{\mu}/p_i\cdot q - p_{j}^{\mu}/p_j\cdot q)$, the one-loop contribution
to the soft current is {\em conserved}:
\begin{equation}
q_\mu J_{a}^{\mu \,(1)}(q, \ep) = 0 \;.
\end{equation}
Combined with the analogous property at tree level, this guarantees that
the soft-gluon factorization formula (\ref{genfac})
is manifestly {\em gauge-invariant}.

The double pole $1/\ep^2$ in Eq.~(\ref{1loopcur}) is the infrared singularity
produced by a soft and collinear virtual gluon. To double-pole accuracy, we can
use colour conservation (see Eq.~(\ref{cocon})) to show that the one-loop
current is simply proportional to the tree-level current:
\beeq
J_{\mu}^{a \,(1)}(q,\ep) \,|\, \cm(q,p_1,\dots,p_m) \rangle
= -\frac{1}{16 \pi^2} \left[ \,\frac{C_A}{\ep^2} J_{\mu}^{a \,(0)}(q)
+ {\cal O}\left(\frac{1}{\ep}\right) \right]
\,|\, \cm(q,p_1,\dots,p_m) \rangle \;.
\eeeq
This behaviour is consistent with the known singularity structure of the
one-loop amplitudes [\ref{GG}, \ref{KS}, \ref{KST}].

Beyond the double-pole approximation, the one-loop current contains
[\ref{Catani:1985dp}] two-particle colour correlations. The correlations
are induced by the last factor on the right-hand side 
of Eq.~(\ref{1loopcur}). This factor fully embodies the logarithmic dependence
on the soft-gluon momentum $q$ and has a simple kinematic interpretation,
being related to the transverse component $q_{\perp, ij}$ of the gluon
momentum with respect to the longitudinal direction singled out by the momenta
$p_i$ and $p_j$ of the colour-correlated hard partons:
\begin{equation}
q^2_{\perp, ij} = \f{2 (p_i\cdot q)\, (p_j\cdot q)}{p_i\cdot p_j} \;.
\end{equation}
The derivation of the result in Eq.~(\ref{1loopcur}) 
(see Sect.~\ref{sec:1loopcur}) suggests that multiparticle colour correlations
will appear in higher-loop contributions to the soft-gluon current 
(\ref{gencur}). For instance, at two-loop order the soft current contains
colour correlations of the type
\begin{equation}
f_{abe} \, f_{ecd} \, T^b_i \, T^c_j \, T^d_k
\end{equation}
between three different hard partons $i, j$ and $k$.

The soft behaviour of the one-loop amplitudes was first investigated in 
Refs.~[\ref{Bern:1995ix}, \ref{Bern:1998sc}, \ref{Bern:1999ry}] by using the
colour-subamplitude formalism. The explicit expressions of the one-loop
soft-gluon contribution to colour subamplitudes, with $m$ external gluons and
with $m$ external gluons plus a $q{\bar q}$ pair,
were derived in Refs.~[\ref{Bern:1998sc}] and [\ref{Bern:1999ry}], 
respectively. The reader can straightforwardly check that the result in 
Eq.~(\ref{1loopcur}) agrees with those in 
Refs.~[\ref{Bern:1998sc}, \ref{Bern:1999ry}] for the particular cases considered
therein. The general result in Eq.~(\ref{1loopcur}) shows that,
although colour and kinematics are deeply entangled in the soft region,
the soft limit of the one-loop amplitudes can be factorized in colour-space in
a way that is both (relatively) simple and process-independent (in particular,
independent of the flavour of the external partons). In particular,
this general structure is quite useful to show (see below and
Sect.~\ref{sec:examples}) 
that colour and kinematics can be completely disentangled in the computation of
the soft limit of the {\em square} of one-loop matrix elements
with two or three external QCD partons.
The factorization formula (\ref{1loopff}) can be used to compute the soft limit
of the one-loop contribution to the square of the matrix element\footnote{In 
the following
the dependence of the matrix element on the
momenta $p_1,\dots,p_m$ of the hard partons is
denoted by $\{p\}$.}
$\cm(q,\{p\})$. Summing over the polarizations of the soft gluon
and
using Eqs.~(\ref{eikfac}) and (\ref{1loopff}),
we have
\beeq
\la&&\!\!\!\!\! \!\!\!\!\!\!\!\!\cm^{(0)}(q,\{p\}) \,| 
\,\cm^{(1)}(q,\{p\}) \,\ra + \;{\rm c.c.} \simeq
- \left( g_S \, \mu^\ep \right)^2 \nn \\
\label{1loopsquare}
&\cdot&\!\!
\Bigg\{ \Big[ \la \,\cm^{(0)}(\{p\}) \,|
\, {\bom J}_{\mu}^{(0)}(q) \cdot {\bom J}^{\mu \,(0)}(q) \,
| \,\cm^{(1)}(\{p\}) \,\ra + \;{\rm c.c.} \Big] \\
&+& \!\!\left( g_S \, \mu^\ep \right)^2 
\Big[ \la \,\cm^{(0)}(\{p\}) \,| \,
{\bom J}_{\mu}^{(0)}(q)  \cdot
{\bom J}^{\mu \,(1)}(q, \ep)
\, | \,\cm^{(0)}(\{p\}) \,\ra + \;{\rm c.c.} \Big] \Bigg\} \;,\nn
\eeeq
where c.c. denotes the complex conjugate. 

The first term on the right-hand 
side, which can be evaluated by using the expression (\ref{eikcur}) of the 
tree-level gluon current, has the same structure as Eq.~(\ref{ccfact}):
\begin{equation}
\label{1looptriv}
\la \,\cm^{(0)}(\{p\}) \,|\, 
{\bom J}_{\mu}^{(0)}(q) \cdot {\bom J}^{\mu \,(0)}(q) \,
| \,\cm^{(1)}(\{p\}) \,\ra + \;{\rm c.c.} = 
2 \,\sum_{i,j=1}^m\, {\cal S}_{ij}(q) \;
|\cm^{(1)}_{(i,j)}(\{p\})|^2 \;,
\end{equation}
where ${\cal S}_{ij}(q)$ is the eikonal function in Eq.~(\ref{eikfun}).
In particular, the two-particle colour correlations 
on the right-hand side
are completely analogous to those at tree level (see Eqs.~(\ref{ccfact})
and (\ref{colam})) and have been taken into account by defining
the colour-correlated one-loop amplitude
\begin{equation}
|\cm^{(1)}_{(i,j)}(\{p\})|^2\equiv\bra{\cm^{(0)}(\{p\})}\;{\bom T}_i \cdot 
{\bom T}_j\,|\cm^{(1)}(\{p\})\rangle + \;{\rm c.c.} \;\;.
\end{equation}

Using expression (\ref{1loopcur}) for the one-loop contribution to the
soft-gluon current, the second term on the right-hand side of 
Eq.~(\ref{1loopsquare}) can be written as
\begin{align}
\la \,\cm^{(0)}(\{p\}) \,| \, {\bom J}_{\mu}^{(0)}(q)  \cdot
&{\bom J}^{\mu \,(1)}(q, \ep) \, | \,\cm^{(0)}(\{p\}) \,\ra + \;{\rm c.c.} =
- \f{1}{4\pi^2}\; \frac{(4\pi)^\ep}{\ep^2} \;\f{\Gamma^3(1-\ep)
\,\Gamma^2(1+\ep)}{\Gamma(1-2\ep)} \nn \\
\label{1loopcolcor}
&\cdot \Bigg\{ C_A \; \cos (\pi \ep) \;{\sum_{i,j}}^{\prime} 
\left[ {\cal S}_{ij}(q) \right]^{1+\ep}
\; | \cm^{(0)}_{(i,j)}(\{p\}) |^2 \\
& + 2 \sin (\pi \ep) \;{\sum_{i,j,k}}^{\prime} {\cal S}_{ki}(q) \,
\left[ {\cal S}_{ij}(q)\right]^{\ep} \left(\lambda_{ij} - \lambda_{iq}
-\lambda_{jq} \right) 
\; | \cm^{(0)}_{(k,i,j)}(\{p\}) |^2 \Bigg\} \nn \;,
\end{align}
where we have used
$\left(\lambda_{ij} - \lambda_{iq}-\lambda_{jq}\right)^2 =1$ for any possible
configuration of incoming and outgoing momenta, and 
the notation $\sum^\prime$ stands for the sum over the different values
of the indices $(i\neq j, j\neq k, k \neq i)$. 

The one-loop contribution on the right-hand side of Eq.~(\ref{1loopcolcor})
contains two terms.
The first term only involves colour correlations between two hard partons,
which are taken into account by the colour-correlated tree-amplitude
$\cm^{(0)}_{(i,j)}(\{p\})$ defined in Eq.~(\ref{colam}).
Thus, apart from an overall $\ep$-dependent factor, its effect simply amounts 
to the rescaling ${\cal S}_{ij}(q) \to \left[ {\cal S}_{ij}(q) \right]^{1+\ep}$ 
in the tree-level factorization formula (\ref{ccfact}). 
The second term instead
has a different structure, because it leads to colour correlations between
three different hard partons. These are included in the
three-parton correlated tree-amplitude:
\begin{equation}
|\cm^{(0)}_{(k,i,j)}(\{p\})|^2 \equiv
f_{abc}\,\langle\cm^{(0)}(\{p\})\,|\;T_k^a\;T_i^b\;T_j^c\;
|\,\cm^{(0)}(\{p\})\,\rangle\, .
\end{equation}

Note that the second term contributes only when there are four or more hard 
partons, because in the case of three partons colour conservation 
(see Eq.~(\ref{cocon})) can be used
to show that the three-parton correlations vanish:
\begin{equation}
f_{abc} T_1^a  T_2^b T_3^c \,|\,\cm^{(0)}(p_1,p_2,p_3)\,\ra =
- f_{abc} T_1^a  T_2^b (T_1^c + T_2^c) \,|\,\cm^{(0)}(p_1,p_2,p_3)\,\ra = 0 \;.
\end{equation}
The absence of these correlations extremely simplifies the structure of the 
soft limit of the squared matrix elements with two or three hard partons.
As a matter of fact, in these cases
any product ${\bom T}_i \cdot {\bom T}_j$ can be expressed as a linear
combination of Casimir operators (see Sect.~\ref{sec:examples}). Therefore,
the colour algebra can explicitly be carried out and the soft limit
of $|\cm(q,\{p\})|^2$ is directly proportional to
$|\cm(\{p\})|^2$ at one-loop accuracy. The corresponding explicit expressions 
are presented in Sect.~\ref{sec:examples}.
 
Note also that in the limit $\ep \to 0$
the three-particle correlation contribution to 
Eq.~(\ref{1loopcolcor}) is at most as singular as $1/\ep$. More precisely,
it gives rise to single poles $1/\ep$ only when there are two or more
incoming partons. In fact, in the case of an outgoing soft gluon $q$, we have 
$\lambda_{ij} - \lambda_{iq}-\lambda_{jq} = +1$ when $i$ and $j$ are both
incoming momenta and $\lambda_{ij} - \lambda_{iq}-\lambda_{jq} = -1$ otherwise.
Therefore, we can rewrite the second term in the curly bracket of 
Eq.~(\ref{1loopcolcor}) as 
\beeq
&&+ 4 \sin (\pi \ep) \;{\sum_{i,j({\rm in}),k}}^{\prime} {\cal S}_{ki}(q) \,
\left[ {\cal S}_{ij}(q)\right]^{\ep} f_{abc} \la \,\cm^{(0)}(\{p\}) \,|
T_k^a  T_i^b T_j^c \,| \,\cm^{(0)}(\{p\}) \,\ra \nn \\
\label{1loopccprime}
&&- 2 \sin (\pi \ep) \;{\sum_{i,j,k}}^{\prime} {\cal S}_{ki}(q) \,
\left[ {\cal S}_{ij}(q)\right]^{\ep} f_{abc} \la \,\cm^{(0)}(\{p\}) \,|
T_k^a  T_i^b T_j^c \,|
\,\cm^{(0)}(\{p\}) \,\ra \;,
\eeeq
where the sum ${\sum^{\prime}_{i,j({\rm in}),k}}$ denotes the restriction of 
${\sum^{\prime}_{i,j,k}}$ to the indices $i,j$ of the incoming partons.
Then, it is easy to show that the second
term in Eq.~(\ref{1loopccprime}) is of ${\cal O}(\ep^2)$ in the limit 
$\ep \to 0$:
\beeq
&&- 2 \sin (\pi \ep) \;{\sum_{i,j,k}}^{\prime} {\cal S}_{ki}(q) \,
\left[ {\cal S}_{ij}(q)\right]^{\ep} f_{abc} \la \,\cm^{(0)}(\{p\}) \,|
T_k^a  T_i^b T_j^c \,|
\,\cm^{(0)}(\{p\}) \,\ra = \\
&&- 2 \pi \ep \;{\sum_{i,j,k}}^{\prime} {\cal S}_{ki}(q) \,
f_{abc} \la \,\cm^{(0)}(\{p\}) \,|
T_k^a  T_i^b T_j^c \,|
\,\cm^{(0)}(\{p\}) \,\ra + {\cal O}(\ep^2) = \nn \\
&&+ 2 \pi \ep \;{\sum_{i,k}}^{\prime} {\cal S}_{ki}(q) \,
f_{abc} \la \,\cm^{(0)}(\{p\}) \,|
T_k^a  T_i^b (T_i^c + T_k^c) \,|
\,\cm^{(0)}(\{p\}) \,\ra + {\cal O}(\ep^2) = {\cal O}(\ep^2) \;.
\eeeq
Here we have set $\left[{\cal S}_{ij}(q)\right]^{\ep} \to 1$. Then we
have used colour conservation, $\sum_{j\neq i,k} T_j^c = - (T_i^c + T_k^c)$,
and the identity $f_{abc} T_k^a  T_i^b (T_i^c + T_k^c) = 0$ for $i\neq k$.

%

\section{Proof of factorization and\\ calculation of the 
one-loop current}
\label{sec:1loopcur}

In this section we derive the factorization formula (\ref{genfac}) 
at one-loop order and we explicitly compute the one-loop contribution to 
the soft-gluon current.

To simplify the analysis, it is convenient to work in a gauge with only 
physical gluon polarizations. We use the axial gauge $n \cdot A=0$ with a 
light-like ($n^2=0$) gauge vector $n^\mu$. 
The polarization vectors $\vep^\mu(k)$ of a gluon with momentum $k$ thus
fulfil the relations $n \cdot \vep(k)=0$ and
$k \cdot \vep(k) \propto k^2$. The sum over the gluon polarizations leads
to the polarization tensor $d^{\mu \nu}$:
\begin{equation}
\label{polten}
d^{\mu \nu}(k) = \sum_{{\rm pol.}} \vep^\mu(k) \; \vep^\nu(k)=
- g^{\mu \nu} + \frac{k^\mu n^\nu + n^\mu k^\nu}{n \cdot k} \;\;.
\end{equation}
The expression on the right-hand side corresponds to
a dimensional-regularization scheme with $d-2=2-2\ep$ gluon polarizations.
For the sake of definiteness we use this explicit expression in all the
intermediate steps of the calculation. However, we shall show that
the final results are regularization-scheme independent by pointing out 
in which steps the scheme dependence might arise.

\subsection{Proof of factorization}

We study the soft behaviour of one-loop matrix elements
by using the eikonal approximation and the soft-gluon insertion
rules\footnote{We use the eikonal approximation both for real and for virtual
soft gluons. In the case of virtual gluons, the eikonal approximation
is not always justified for each {\em single} Feynman diagram. Nonetheless, it
is valid for any gauge-invariant set of Feynman diagrams, such as those
computed in this section. This can be shown [\ref{Catani:1986xt}]
by simply using time-ordered perturbation theory, where the eikonal approximation is
valid on a graph-by-graph basis.} recalled
above Eq.~(\ref{eikcons}). For this purpose, it is useful to decompose the
one-loop matrix element $\cm^{(1)}$ in three contributions,
\begin{equation}
\label{virtdecom}
\cm^{(1)} = \cm_{hard}^{(1)} + \cm_{coll}^{(1)} + \cm_{soft}^{(1)} \, \;, 
\end{equation}
which respectively represent the kinematic regions where the momentum $k$
circulating in the loop is 
\begin{itemize}
\item {\em hard}: its momentum components are of the same
order as those of the hard external momenta $p$,
\item {\em collinear}: $k$ is parallel to one of the hard external 
momenta,
\item {\em soft}:  its momentum components are much smaller than those
of the hard external momenta $p$.
\end{itemize}

\begin{figure}[htb]
\begin{center}
\begin{tabular}{c}
\epsfxsize=10truecm
\epsffile{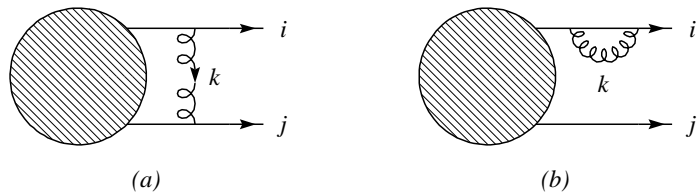}\\
\end{tabular}
\end{center}
\caption{\label{mloop}{\em Feynman diagrams that contribute to the soft
behaviour of the one-loop amplitude. The shaded blob denotes the tree-level
amplitude and the virtual gluon with soft momentum $k$ 
can either $(a)$ connect two different external 
legs $i$ and $j$ or $(b)$ be emitted and reabsorbed by the same leg. 
}}
\end{figure}

For the purpose of the following discussion, we recall that the
soft virtual behaviour of the one-loop matrix element
$\cm^{(1)}(\{p\})$ can be computed by using the soft-gluon insertion rules.
The soft virtual gluon of momentum $k$ is inserted (emitted and reabsorbed)
on all the external legs (Fig.~\ref{mloop}) of the tree-level amplitude 
$\cm^{(0)}(\{p\})$, leading to the expression 
[\ref{GG}--\ref{Catani:1985dp}, \ref{CSdipolelet}]:
\begin{equation}
\label{softvirt}
| \, \cm_{soft}^{(1)}(\{p\}) \, \ra = \frac{1}{2} \;\gs^2 \,\mu^{2\ep}
\int \frac{d^dk}{(2\pi)^d} \;\frac{i}{k^2 + i0} 
\;\left[ \,J_{\mu}^{a \,(0)}(k) \,\right]^\dagger \; J^{\mu \,a \,(0)}(k) \;
| \, \cm^{(0)}(\{p\}) \, \ra \;\;,
\end{equation}
where $J_{\mu}^{a \,(0)}(k)$ is the (tree-level) soft-gluon current in
Eq.~(\ref{eikcur}).

We now consider the one-loop matrix element $\cm^{(1)}(q,\{p\})$
when the momentum $q$ of the external gluon becomes soft. To apply
the soft-gluon insertion rules, we perform the decomposition in 
Eq.~(\ref{virtdecom}) and discuss the three different kinematic regions
of the virtual momentum $k$ in turn.

When $k$ is in the hard region, all the internal lines in $\cm^{(1)}(q,\{p\})$
are highly off-shell. Thus the real soft gluon $q$
can couple only to the external legs and we can factorize its contribution
as in the case of tree-level amplitudes. Neglecting terms that are not
singular in the soft limit $q \to 0$, we get
\begin{equation}
\label{eqhard}
\la a | \,\cm^{(1)}_{hard}(q,\{p\}) \ra \simeq g_S \mu^\ep \vep^\mu(q)
J_\mu^{a \, (0)}(q) \;|\,\cm_{hard}^{(1)}(\{p\}) \ra \, ,
\end{equation} 
where $J_\mu^{a \, (0)}(q)$ is the soft-gluon current in Eq.~(\ref{eikcur}).

We now consider the region where the loop momentum $k$ is collinear to the
momentum of one, say $p_i$, of the hard external legs. 
Since we work in a physical gauge, the only diagram that is not dynamically
suppressed [\ref{BCM}] is that in which the loop leads to a 
self-energy contribution on the external leg $p_i$ 
(e.g. the diagram in Fig.~\ref{mloop}~$(b)$). The soft gluon
can then
be inserted in this diagram in all possible ways: on the line
$p_i$ before and after the self-energy contribution and on the
self-energy lines themselves. However, we can exploit
the {\em colour coherence} properties [\ref{BCM}] of QCD radiation.
Since the loop-momentum $k$ is parallel
to $p_i$, the soft gluon $q$ cannot distinguish the self-energy lines
from a single line with total colour charge $T^a_i$.
The {\em sum} of the
insertions of $q$ on this diagram is thus insensitive to the presence of
the collinear loop momentum and leads (see, for instance, Sect.~3.4 in
Ref.~[\ref{Catani:2000ss}] for a detailed similar discussion) to the same
factor, $T^a_i p_i\cdot \vep(q)/p_i \cdot q$, as in the soft-gluon
factorization at tree level. Considering the insertions of $q$ on all the
other external legs $j \neq i$, we obtain a factorization formula
of tree-level type also in the collinear region:
\begin{equation}
\label{eqcoll}
\la a | \,\cm^{(1)}_{coll}(q,\{p\}) \ra \simeq g_S \mu^\ep \vep^\mu(q)
J_\mu^{a \, (0)}(q) \;|\,\cm_{coll}^{(1)}(\{p\}) \ra \, .
\end{equation} 

We finally have to deal with the region in which the loop momentum $k$
is carried by a soft gluon\footnote{Quarks loops are dynamically 
suppressed when their momentum become soft. Without loss of infrared accuracy,
we thus consider
the quark loops as included in the hard or collinear regions.}.
Unlike the case of the hard and collinear regions, where the 
one-loop effects can be factorized with respect to the tree-level
current $J_\mu^{a \, (0)}(q)$ (see Eqs.~(\ref{eqhard}) and (\ref{eqcoll})),
new `non-factorizable' contributions appear when the loop momentum is soft.
To single out these new contributions, we write the following identity:
\begin{align}
\label{eqsoft}
|\cm^{(1)}_{soft}(q,\{p\})\ra
&= g_S \,\mu^\ep \,\vep^\mu(q) \,{\bom J}^{(0)}_\mu(q) \,
|\cm_{soft}^{(1)}(\{p\}) \ra\nn\\
&+\Big(\, |\cm_{soft}^{(1)}(q,\{p\})\ra
-g_S\mu^\ep\vep^\mu(q) {\bom J}^{(0)}_\mu(q)
|\cm_{soft}^{(1)}(\{p\})\ra\,\Big) \;\;,
\end{align}
where we have added and subtracted the `factorized' contribution. 
Then we combine the contributions from the hard, collinear and soft regions
by adding Eqs.~(\ref{eqhard}), (\ref{eqcoll}) and (\ref{eqsoft}), and we
obtain
\begin{align}
\label{eqtot}
|\cm^{(1)}(q,\{p\})\ra
&= g_S \,\mu^\ep \,\vep^\mu(q) \,{\bom J}^{(0)}_\mu(q) \,
|\cm^{(1)}(\{p\}) \ra\nn\\
&+\Big(\, |\cm_{soft}^{(1)}(q,\{p\})\ra
-g_S\mu^\ep\vep^\mu(q) {\bom J}^{(0)}_\mu(q)
|\cm_{soft}^{(1)}(\{p\})\ra\,\Big) \;\;.
\end{align}
The first term on the right-hand side of Eq.~(\ref{eqsoft})
together with the contributions from Eqs.~(\ref{eqhard}) and (\ref{eqcoll})
have reconstructed the first term on the right-hand side of Eq.~(\ref{eqtot}),
which is exactly the first term on the right-hand side of the factorization
formula (\ref{1loopff}). What remains to be done to prove the factorization
formula is to relate the second term on the right-hand side of 
Eq.~(\ref{1loopff}) with 
the contribution in the round bracket of Eq.~(\ref{eqtot}).

\begin{figure}[htb]
\begin{center}
\begin{tabular}{c}
\epsfxsize=15.8truecm
\epsffile{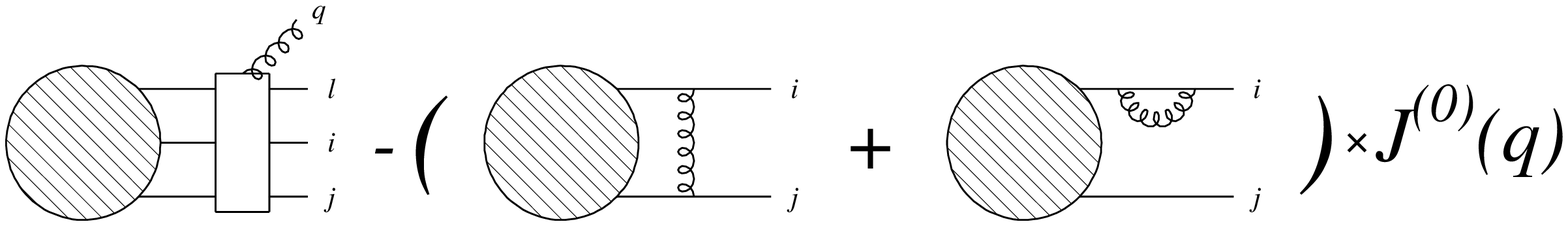}\\
\end{tabular}
\end{center}
\caption{\label{relevant}{\em Graphs that contribute to the 
one-loop soft current.}}
\end{figure}

For this purpose, we first note that when the real gluon $q$ and the virtual
gluon $k$ are both soft, they can couple only to the external hard lines. 
In the corresponding Feynman diagrams, which are schematically represented by
the first graph in Fig.~\ref{relevant}, the tree-level amplitude 
$\cm^{(0)}(\{p\})$ is factorized in the soft limit. We can write:
\begin{equation}
\label{kernel}
|\cm^{(1)}_{soft}(q,\{p\})\ra \simeq \left( g_S \,\mu^\ep \right)^3
\,\vep^\mu(q) \,{\bom K}^{(1)}_\mu(q,\ep) \;|\cm^{(0)}(\{p\})\ra \;\;,
\end{equation}
where the kernel ${\bom K}^{(1)}$ (represented by the box in 
Fig.~\ref{relevant}) denotes all the soft-gluon insertions of 
$q$ and $k$ on the hard-momentum lines.
Then, we note that $\cm^{(0)}(\{p\})$ is factorized also in the expression
(\ref{softvirt}) for $\cm_{soft}^{(1)}(\{p\})$. Therefore, 
the contribution in the round bracket of Eq.~(\ref{eqtot}) can be recast in the
form of the second term on the right-hand side of the factorization formula
(\ref{1loopff}).
Moreover, using Eqs.~(\ref{kernel}) and (\ref{softvirt}), we obtain the
following explicit representation of the one-loop contribution 
${\bom J}^{(1)}$ to the soft-gluon current (Fig.~\ref{relevant}): 
\begin{equation}
\label{1loopk}
\vep^\mu(q) \;{\bom J}_\mu^{(1)}(q,\ep) =
\vep^\mu(q) \left\{ \, {\bom K}^{(1)}_\mu(q,\ep) - 
{\bom J}^{(0)}_\mu(q) \;\frac{1}{2} \;
\int \frac{d^dk}{(2\pi)^d} \;\frac{i}{k^2 + i0} 
\;\left[ \,{\bom J}_{\nu}^{(0)}(k) \,\right]^\dagger \cdot
\; {\bom J}^{\nu \,(0)}(k) \right\} \;.
\end{equation}

\subsection{Calculation of the one-loop current}

\begin{figure}[htb]
\begin{center}
\begin{tabular}{c}
\epsfxsize=15.8truecm
\epsffile{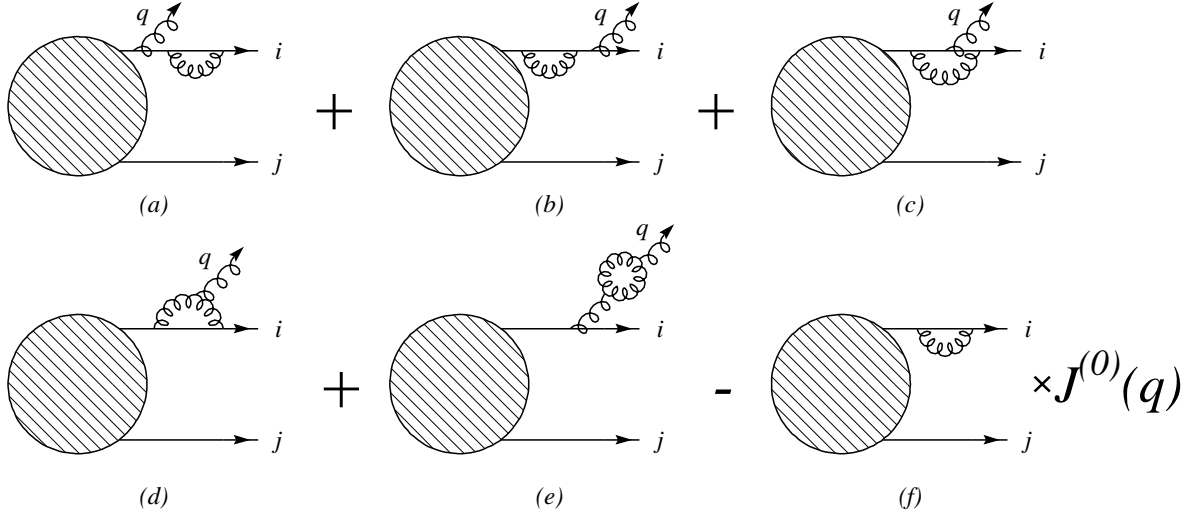}\\
\end{tabular}
\end{center}
\caption{\label{oneleg}{\em Feynman diagrams that depend on a single hard
momentum.
}}
\end{figure}

\begin{figure}[htb]
\begin{center}
\begin{tabular}{c}
\epsfxsize=15truecm
\epsffile{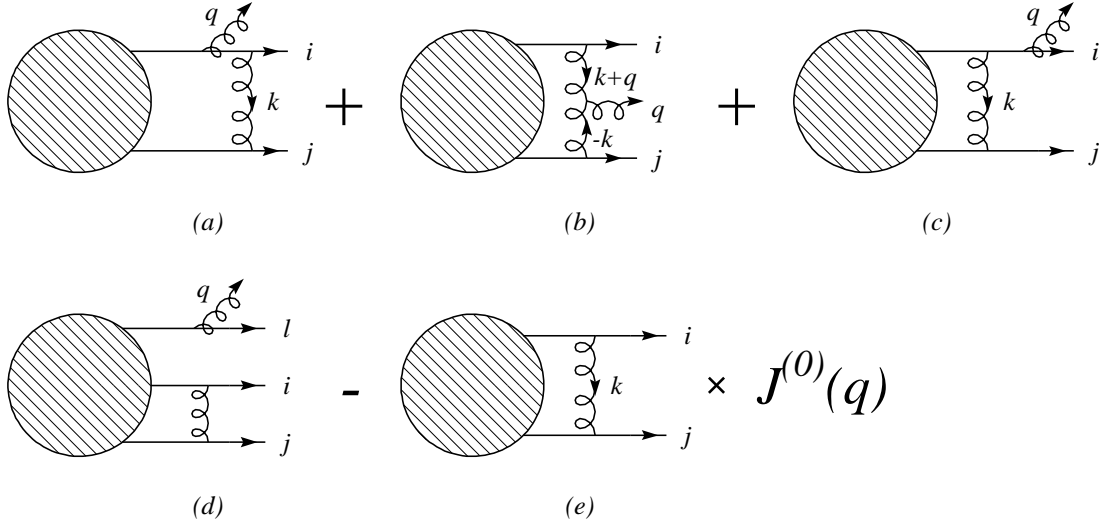}\\
\end{tabular}
\end{center}
\caption{\label{twolegs}{\em Feynman diagrams that depend on two or three
different hard momenta.
}}
\end{figure}

We now proceed to the explicit calculation of the one-loop soft current 
${\bom J}^{(1)}$. Using the eikonal
approximation, we have to evaluate the Feynman diagrams of the kernel
${\bom K}^{(1)}$ and to subtract those corresponding to the second term
on the right-hand side of Eq.~(\ref{1loopk}). We divide the diagrams in two
classes: $(A)$ the diagrams that involve interactions with a single 
hard line (Fig.~\ref{oneleg}), and  $(B)$ all the remaining diagrams
(Fig.~\ref{twolegs}).


We first consider class $(B)$. The diagrams $(a), (b), (c)$ and $(d)$
in Fig.~\ref{twolegs} come from the kernel ${\bom K}^{(1)}$, while 
Fig.~\ref{twolegs}~$(e)$ represents the corresponding subtractions.  We see 
that within this class there are diagrams that involve
interactions between three different external lines $i,j$ and $l$. These
diagrams cancel.
Indeed, the diagram in 
Fig.~\ref{twolegs}~$(d)$ is exactly cancelled by that $(e)$-contribution
in which $q$ is emitted by the line $l$ (the sum over the emissions of $q$ 
is included in the factor ${\bom J}^{(0)}(q)$). Thus the only non-vanishing
terms coming from class $(B)$ are those that involve interactions between
two external lines, namely the diagrams $(a), (b), (c)$ and the corresponding
subtractions in $(e)$. The diagrams $(a), (c)$ and the subtractions in $(e)$
are very similar: they can be combined in a simple way,
because they only
differ by the momentum and colour flow along the hard line $i$. The different
factors coming from the line $i$ in the diagrams $(c)$ and $(e)$ give
\begin{equation}
\label{cediag}
+ T_i^a T_i^b \;\frac{1}{p_i q +i0} \;\frac{1}{p_i (k+q) +i0}
- T_i^a T_i^b \;\frac{1}{p_i q +i0} \;\frac{1}{p_i k +i0} \; ,
\end{equation}
while the corresponding factor from the diagram $(a)$ is
\begin{equation}
\label{adiag}
+ T_i^b T_i^a \;\frac{1}{p_i k +i0} \;\frac{1}{p_i (k+q) +i0} \;.
\end{equation}
Decomposing the colour factor in Eq.~(\ref{adiag}) in its non-abelian
and abelian components,
$T_i^b T_i^a = i f_{bac} T_i^c + T_i^a T_i^b$, and adding Eq.~(\ref{cediag}), 
we see that the abelian component $T_i^a T_i^b$ of the
diagram $(a)$ exactly cancels the diagrams $(c)$ and $(e)$. In conclusion,
the total contribution to the one-loop current from class $(B)$ simply
amounts to computing the diagram $(b)$, which is non-abelian, and the 
non-abelian part of the diagram $(a)$, which is obtained by the 
replacement $T_j^b T_i^b T_i^a \to i f_{bac} T_j^b T_i^c$ ($b$ is the colour
index of the virtual gluon) in its overall colour factor.

The calculation of the diagrams in Fig.~\ref{twolegs}~$(a)$ and $(b)$ is quite
simple (see below). Then we have to add the diagrams of class $(A)$ 
(Fig.~\ref{oneleg}). An argument similar to that in Eqs.~(\ref{cediag}) and 
(\ref{adiag}) can be used to show that the abelian parts of these diagrams
cancel\footnote{More precisely, the abelian diagrams in 
Figs.~\ref{oneleg}~$(a)$ and $(b)$ are cancelled by the abelian part of the
diagram in Fig.~\ref{oneleg}~$(c)$.}. Thus, only their non-abelian part
has to be evaluated. Although straightforward, this calculation is 
quite cumbersome, in particular because it has to be carried out in the axial
gauge $n\cdot A=0$. This cumbersome calculation can be short-circuited by
exploiting gauge invariance.

We first recall that the non-vanishing contributions from class $(B)$
depend on the momenta and charges of (at most) two hard partons.
Since the diagrams of class $(A)$ involve interactions with
a single hard line, we can split the
one-loop soft current in two terms, $J_{1P}$ and $J_{2P}$, that
respectively denote the contributions that depend on one and two hard
momenta:
\begin{equation}
\label{cursplit}
J_{\mu}^{a(1)}(q,\ep)=J_{\mu;1P}^{a(1)}(q,\ep)+J_{\mu;2P}^{a(1)}(q,\ep)\, .
\end{equation}
All the diagrams of class $(A)$ are included in $J_{1P}$. Those of class
$(B)$ contribute to $J_{2P}$ and (because of possible
cancellations in the dependence of one hard momentum) to $J_{1P}$. We shall
explicitly compute $J_{2P}$ and show that it is gauge-invariant. Thus
$J_{1P}$ has to be gauge invariant as well, that is, it cannot depend
on the gauge vector $n^\nu$. Power counting and the fact that the one-loop
current is non-abelian are then sufficient to determine
$J_{1P}$, apart from an overall factor $f^{(1)}(\ep)$ that can only
depends on~$\ep$:
\begin{equation}
J_{\mu;1P}^{a \,(1)}(q,\ep)= C_A \;f^{(1)}(\ep) \sum_{i=1}^m (p_i \cdot q)^{-\ep}
\f{p_{i\mu}}{p_i\cdot q}\;T^a_i\; .
\end{equation}
This overall factor can then be obtained by imposing that the soft-gluon
factorization formula be gauge-invariant (with respect
to the polarizations of the real gluon $q$) or, equivalently, 
that the complete one-loop current in Eq.~(\ref{cursplit}) be conserved.
This leads to the constraint
\begin{equation}
\label{conscosnt}
q^\mu J_{\mu}^{a(1)}(q,\ep) = q^\mu J_{\mu;2P}^{a \,(1)}(q,\ep) + C_A \;
f^{(1)}(\ep) \sum_{i=1}^m (p_i \cdot q)^{-\ep} \;T^a_i = 0 \;\;,
\end{equation}
which can be used to get $f^{(1)}(\ep)$ (and, hence, $J_{\mu;1P}^{a(1)}$)
from the two-parton contribution $J_{\mu;2P}^{a(1)}$. In particular, the
calculation of the diagrams of class $(A)$ can be avoided.

We anticipate that our explicit calculation of the two-parton 
contribution $J_{\mu;2P}^{a(1)}$ gives $q^\mu J_{\mu;2P}^{a \,(1)}(q,\ep) = 0$.
Thus the single-parton contribution $J_{\mu;1P}^{a(1)}$ vanishes.
Note that this does not mean that the diagrams of class $(A)$ give
a vanishing contribution, but rather that their contribution is cancelled
by the terms of class $(B)$ that depend on a single hard momentum. 

We can now complete our calculation of the one-loop current by
explicitly computing the non-abelian part of the diagrams in 
Figs.~\ref{twolegs}~$(a)$ and $(b)$, and, more precisely, their contribution
to $J_{\mu;2P}^{a(1)}$. For the sake of definiteness, we 
write down the diagrams for the case in which all external momenta are 
outgoing. The diagram in Fig.~\ref{twolegs}~$(a)$ depends on the polarization
tensor $d(k)$ of the virtual gluon. Using the expression in Eq.~(\ref{polten})
and performing the Lorentz algebra of the spin numerator, 
the non-abelian part of the diagram in Fig.~\ref{twolegs}~$(a)$ gives
\begin{align}
\label{commu}
  f_{abc}\sum_{i\neq
  j}\,T_i^c \,T^b_j \int &\f{d^dk}{(2\pi)^d}\; \f{1}{(k^2+i0)(p_j\cdot
  k-i0)}\; \f{p_i \cdot \vep(q)}{p_i\cdot(k+q)+i0} \nn \\
  & \cdot \f{1}{p_i\cdot k+i0} \;
  \Bigg( p_i\cdot p_j - p_j\cdot k \; \f{p_i\cdot n}{k\cdot n} 
  - p_i\cdot k \;\f{p_j\cdot n}{k\cdot n} \Bigg) \;.
\end{align}
The diagram\footnote{Note that we use the eikonal approximation
for the vertices and propagators of the hard-momentum lines $i$ and $j$,
while gluon propagators and the three-gluon vertex must be treated exactly.} 
in Fig.~\ref{twolegs}~$(b)$ depends on the polarization tensors $d(k)$ and
$d(k+q)$ of the virtual-gluon lines. Performing the Lorentz algebra of the spin 
numerator and exploiting the symmetry of the integrand under the transformation 
$i\lra j, k\lra -(k+q)$, the diagram in Fig.~\ref{twolegs}~$(b)$ gives
\begin{align}
\label{nab}
&f_{abc}\sum_{i\neq j} T_i^c T^b_j \int\f{d^dk}{(2\pi)^d}
\; \f{1}{(k^2+i0)(p_j\cdot k-i0)}\; \f{1}{p_i\cdot(k+q)+i0} \;
\f{1}{(k+q)^2+i0} \;\vep_\mu(q) \nn \\
&\cdot 
\Bigg[ k^\mu\, p_i\cdot p_j- p_i^\mu \, p_j\cdot (k+2q) - p_j\cdot k
\left( 2 k^\mu \f{p_i\cdot n}{k\cdot n} - p_i^\mu 
\f{(k+2q)\cdot n}{k\cdot n} \right) 
 + p_i^\mu (k+q)^2 \;\f{p_j\cdot n}{k\cdot n}\Bigg] \;.
\end{align}
By inspection of Eqs.~(\ref{commu}) and (\ref{nab}), we see that the last term
in the round bracket of Eq.~(\ref{commu}) cancels the last term  
in the square bracket of Eq.~(\ref{nab}). The remaining $n$-dependence
of the integrands is proportional to $p_j\cdot k$. This factor cancels
the fermion propagator $1/(p_j\cdot k-i0)$, thus leaving a contribution that,
although $n$-dependent, depends on a single hard momentum $p_i$. This 
term can then be included, together with the class $(A)$ diagrams,
in the single-particle contribution $J_{\mu;1P}^{(1)}(q,\ep)$ to the one-loop
current. We conclude that 
(the contribution of Eqs.~(\ref{commu}) and (\ref{nab}) to)
the two-particle current $J_{\mu;2P}^{(1)}(q,\ep)$ is explicitly independent
of the gauge vector $n^\mu$. Note also that $n$-dependence cancels at the
integrand level, and thus, the cancellation is completely insensitive to the
actual prescription [\ref{bassetto}] 
used to regularize the gauge pole $1/(n\cdot k)$ of the
gluon polarization tensor (\ref{polten}). 

Although the expression (\ref{polten}), which we have used for the 
polarization tensor, corresponds to $d-2=2-2\ep$ helicity states for the gluon,
our calculation also allows us to discuss the (in)dependence on the 
dimensional-regularization scheme. Other regularization schemes 
[\ref{Sie79}, \ref{4dhs}] use 2 helicity states for the gluon. At
one-loop order, the difference eventually amounts [\ref{KST2to2}, \ref{uni}] to 
set $\ep=0$ in the {\em integrands} after having performed the Lorentz algebra
of the spin numerators. Since the integrands of Eqs.~(\ref{commu}) and 
(\ref{nab}) have no explicit $\ep$-dependence, our result for 
$\vep(q) \cdot J_{2P}^{(1)}$ (and hence for the complete current
$\vep(q) \cdot J^{(1)}$) does not depend on the scheme used to implement
dimensional regularization.

We proceed to the computation of  $\vep(q) \cdot J_{2P}^{(1)}$ by adding
the two-particle contributions from Eqs.~(\ref{commu}) and (\ref{nab}).
As for Eq.~(\ref{nab}), the term proportional to $k^\mu$  can be reduced to 
scalar integrals. Then using $q \cdot\vep(q)= 0$ and exploiting the symmetry 
with respect to the transformation  $i \lra j, k \lra -(k+q)$,
we obtain
\begin{align}
\label{inte}
\vep(q) \cdot J_{2P}^{a \,(1)}(q,\ep)
&= f_{abc}\,\sum_{i\neq j} T_i^c\;
T^b_j\; \int\f{d^dk}{(2\pi)^d} \;\f{p_i\cdot\vep(q)}{p_i\cdot(k+q)+i0}\,
\,\f{1}{(k^2+i0) (p_j\cdot k-i0)} \nn\\
&\cdot\;\Bigg[ \;\f{p_i\cdot p_j}{p_i\cdot k+i0}+\f{1}{(k+q)^2+i0}
\,\left( \f{p_i\cdot p_j}{p_i\cdot q}\, k\cdot q-2p_j\cdot q
-\f{p_j\cdot q}{p_i\cdot q}\, p_i\cdot k\right)\Bigg].
\end{align}
The ultraviolet-finite integral can be easily performed in the collinear frame
in which $p_i$ and $p_j$ are directed along the `$+$' and `$-$' light-cone
directions, respectively. In this frame, the integration over the $k_+$-complex
plane receives contributions from the pole $1/(p_j\cdot k-i0)$ in the fermion 
propagators and from the poles in the gluon propagators. Closing the integration
contour on the lower half-plane, and using the residue theorem, we can select 
only the gluon poles, which amounts to the following replacements in
Eq.~(\ref{inte}):
\begin{equation}
\frac{1}{k^2+i0} \to -2 \pi i \,\delta_+(k^2) \;, \;\;\;\;
\frac{1}{(k+q)^2+i0} \to -2 \pi i \,\delta_+((k+q)^2) \;\;.
\end{equation}
Therefore Eq.~(\ref{inte}) can be expressed as follows
\begin{align}
\vep(q) \cdot J_{2P}^{a \,(1)}(q,\ep)&=-\f{i}{2}\, 
f_{abc}\sum_{i\neq j}T_i^c\;
T^b_j\,\f{p_{i}\cdot\vep(q)}{p_i\cdot q}\Big[I(p_i,p_j;p_i;2p_i\cdot q)
+I(p_i,p_j;p_j;-2p_j\cdot q-i0)\nn\\
\label{masintex}
&-I(q,p_j;p_j;-2p_j\cdot q-i0)-I(p_j,q;p_i;2p_i\cdot q)
- I(p_i,q;p_j;-2p_j\cdot q-i0)\Big] \;,
\end{align}
in terms of the master integral
\begin{equation}
\label{masint}
I(p,\pb;r;s) = \int\f{d^dk}{(2\pi)^{d-1}} \;\delta_+(k^2)
\; \f{p\cdot\pb}{(p\cdot k)\,(\pb\cdot k)} \;\f{s}{2r\cdot k+s} \;\;,
\end{equation}
which depends on the light-like momenta $p^\mu,\pb^\mu$ and $r^\mu$ and on 
the complex scalar $s$.
The evaluation of the $d$-dimensional integration in Eq.~(\ref{masint})
over the on-shell gluon momentum $k$ is straightforward and gives
\begin{equation}
\label{intres}
I(p,\pb;r;s) =\f{1}{8\pi^2}\left[ 
\f{8\pi (p\cdot r)\,(\pb \cdot r)}{s^2 \,(p\cdot\pb)}\right]^\ep\;\f{1}{\ep^2}
\;\Gamma^2(1-\ep)\;\Gamma(1+\ep)\;\Gamma(1+2\ep) \;\;.
\end{equation}
Since $I(p,\pb;r;s)$ vanishes when $r=p$ or $r=\pb$, inserting
Eq.~(\ref{intres}) in Eq.~(\ref{masintex}), we finally obtain
\begin{align}
\vep(q) \cdot J_{2P}^{a \,(1)}(q,\ep) &=\f{1}{16\pi^2}\;  
i\; f_{abc}\sum_{i\neq j}T_i^c\;
T^b_j \;\f{p_{i}\cdot\vep(q)}{p_i\cdot q}\left(\f{4\pi\,
  p_i\cdot p_j}{2p_i\cdot q\, p_j\cdot
  q}\right)^\ep \nn\\
&~\cdot \f{1}{\ep^2}\,\Gamma^2(1-\ep)\Gamma(1+\ep)\Gamma(1+2\ep)
\Bigg[1+\left(\f{1}{e^{-2\pi i}}\right)^\ep\Bigg]\nn\\
&=-\f{1}{16\pi^2}\;  i\; f_{abc}\sum_{i\neq j}T_i^b\;
T^c_j \; \vep_{\mu}(q)\left(\f{p_{i}^\mu}{p_i\cdot q}-\f{p_{j}^\mu}
{p_j\cdot q}\right)\left(\f{4\pi\,
  p_i\cdot p_j}{2p_i\cdot q\, p_j\cdot
  q\,e^{-i\pi}}\right)^\ep \nn\\
\label{1loopfin}  
&~\cdot \f{1}{\ep^2}\f{\Gamma^3(1-\ep)\,\Gamma^2(1+\ep)}{\Gamma(1-2\ep)}\, ,
\end{align}
where we have used the identity
\begin{equation}
\label{identity}
\cos (\pi \ep) = 
\frac{\Gamma(1+\ep) \,\Gamma(1-\ep)}{\Gamma(1+2\ep) \,\Gamma(1-2\ep)} \;,
\end{equation}
and the symmetry under the replacement $i \lra j$.

Note that the current in Eq.~(\ref{1loopfin}) is conserved, 
$q^\mu J_{\mu;2P}^{a \,(1)}(q,\ep)=0$. Therefore, as discussed below
Eq.~(\ref{conscosnt}), the single-particle term $J_{\mu;1P}^{a \,(1)}$
vanishes and Eq.~(\ref{1loopfin}) gives the complete one-loop contribution
to the soft-gluon current. We conclude that we have explicitly derived the
result in Eq.~(\ref{1loopcur}) in the case of outgoing parton momenta 
$p_i, p_j$ (when $\lambda_{ij}=\lambda_{iq}= \lambda_{jq}=+1).$ The completely
general result in Eq.~(\ref{1loopcur}) is straightforwardly obtained
by crossing ($p_i \to - p_i$ and/or $p_j \to - p_j$) the parton momenta in
Eq.~(\ref{inte}) and repeating the steps that lead to Eq.~(\ref{1loopfin}).

\section{Processes with two and three partons \\
and \ee $\to 3$~jets at NNLO}
\label{sec:examples}

In general soft factorization formulae involve colour correlations
(see Eqs.~(\ref{ccfact}) and (\ref{1loopcolcor})). As discussed at the end
of Sect.~\ref{sec:1loopfac}, in the case of processes with two or three hard 
external partons, the correlations are completely given in terms of the products
of the colour-charge factors ${\bom T}_i\cdot{\bom T}_j$. In this case,
moreover, these colour-charge factors can be expressed in terms of the
Casimir operators of the hard partons. Therefore the colour algebra can 
explicitly be carried out and we can obtain QED-like factorization formulae.
We first recall the relations between colour-charge factors and Casimir
operators and then give the factorization formulae. 

When there are only two hard partons in the amplitude, we can use colour
conservation $({\bom T}_1= -{\bom T}_2)$ to obtain
\begin{equation}
\label{cas2}
{\bom T}_1\cdot{\bom T}_2 \;\ket{\cm(p_1,p_2)}=
-{\bom T}_1^2 \;\ket{\cm(p_1,p_2)}= - C_1 \;\ket{\cm(p_1,p_2)}
= - C_2 \;\ket{\cm(p_1,p_2)}\, ,
\end{equation}
where $C_1=C_2$ is the Casimir of the hard partons.
When the hard partons are three, the colour algebra can still be performed
in closed form. Let us consider, for instance, 
the correlation term ${\bom T}_1\cdot{\bom
  T}_2$. Using colour conservation 
$({\bom T}_3= -{\bom T}_1-{\bom T}_2)$, 
its action on the ket $\ket{\cm(p_1,p_2,p_3)}$ is
\begin{equation}
\label{cas3}
2 {\bom T}_1\cdot{\bom T}_2\ket{\cm(p_1,p_2,p_3)}=({\bom
  T}_3^2-{\bom T}_1^2-{\bom T}_2^2)\ket{\cm(p_1,p_2,p_3)}=
  (C_3-C_1-C_2)\ket{\cm(p_1,p_2,p_3)}
\end{equation}
and thus ${\bom T}_1\cdot{\bom T}_2$ is again given in terms of the
Casimir invariants $C_i$. 
The same can be done for the other products ${\bom T}_1\cdot
{\bom T}_3$ and ${\bom T}_2\cdot{\bom T}_3$.

When the number of hard partons is four or more, colour correlations cannot be
avoided.
In fact the number of independent equations following from colour conservation 
is not sufficient to express the products ${\bom T}_i\cdot{\bom T}_j$ in terms of
Casimir invariants.

The colour-algebra results in Eqs.~(\ref{cas2}) and (\ref{cas3}) can be 
inserted in Eqs.~(\ref{ccfact}), (\ref{1looptriv}) and (\ref{1loopcolcor})
to straightforwardly obtain the soft limit of the squared amplitudes
with a soft gluon and two or three hard partons (plus any number of
colourless partons). In both cases, the contributions from 
Eqs.~(\ref{ccfact}) and (\ref{1loopsquare}) reconstruct the one-loop
expansion of the all-order squared amplitude $|\cm(\{p\})|^2$ in 
Eq.~(\ref{loopexp}),
and we obtain the following factorization formulae
\begin{equation}
\label{2part}
| \cm(q,p_1,p_2) |^2 \simeq
\gs^2 \,\mu^{2\ep} \,4 \,C_1 \, {\cal I}_{12}(q) 
\;| \cm(p_1,p_2) |^2 \;\;,
\end{equation}
\begin{equation}
\label{3part}
| \cm(q,p_1,p_2,p_3) |^2 \simeq \gs^2 \,\mu^{2\ep} \,2 \,
\Big[ 2 C_1 {\cal I}_{12}(q) + C_A \Big( {\cal I}_{23}(q) + 
{\cal I}_{13}(q) - {\cal I}_{12}(q) \Big) \Big]
 \;| \cm(p_1,p_2,p_3) |^2 \;\;,
\end{equation}
where the kinematic factor ${\cal I}_{ij}(q)$ includes the loop corrections to
the eikonal function ${\cal S}_{ij}(q)$ in Eq.~(\ref{eikfun}). To one-loop
accuracy, we find
\begin{equation}
{\cal I}_{ij}(q) = {\cal S}_{ij}(q) \cdot \Big\{ 1 - C_A \,\f{\gs^2}{8\pi^2}
\;\f{1}{\ep^2} 
\;\f{\Gamma^4(1-\ep)\,\Gamma^3(1+\ep)}{\Gamma^2(1-2\ep)\,\Gamma(1+2\ep)}
\;\Big[ 4 \pi \mu^2 {\cal S}_{ij}(q) \Big]^{\ep} + {\cal O}(\gs^4)
\Big\} \;\;,
\end{equation}
where we have used Eq.~(\ref{identity}).
Note that in the amplitude with three hard partons, two of them
(say, the ones with momenta $p_1$ and $p_2$) have to form a
particle--antiparticle pair (either a $q{\bar q}$ pair or a gluon pair)
and the third one (with momentum $p_3$) 
has to be a gluon. In Eq.~(\ref{3part}) we have therefore set $C_1=C_2$ and
$C_3=C_A$.

The results in Eqs.~(\ref{2part}) and (\ref{3part}), when combined with the 
analogous QED-like factorization formulae for the emission of two soft gluons 
[\ref{glover}, \ref{Catani:2000ss}] and of two [\ref{Kosower:1999rx}, 
\ref{Bern:1999ry}] and three [\ref{glover}, \ref{Catani:2000ss}] 
collinear partons, can be used to perform cross section calculations at NNLO 
for several processes. For instance, Eq.~(\ref{2part}) is relevant to
2-jet production in $e^+e^-$ annihilation and for the production of Drell--Yan,
photon and vector-boson pairs in hadron collisions, while Eq.~(\ref{3part})
can be used for the production of 3 jets in $e^+e^-$ annihilation, (2+1) jets
in deep-inelastic lepton--hadron collisions and (vector boson + jet) in 
hadron collisions.
Owing to the high precision of the data from $e^+e^-$-collider experiments,
improved theoretical calculations for 3-jet production are highly demanded.
The fact that colour and kinematics factors are completely disentangled in the
singular limits of the corresponding matrix elements certainly simplifies the
structure of these calculations.

\noindent {\bf Acknowledgements.}\\
\noindent
We would like to thank Vittorio Del Duca and Zoltan Kunszt for comments. 

\section*{References}

\def\ac#1#2#3{Acta Phys.\ Polon.\ #1 (19#3) #2}
\def\ap#1#2#3{Ann.\ Phys.\ (NY) #1 (19#3) #2}
\def\ar#1#2#3{Annu.\ Rev.\ Nucl.\ Part.\ Sci.\ #1 (19#3) #2}
\def\cpc#1#2#3{Computer Phys.\ Comm.\ #1 (19#3) #2}
\def\ib#1#2#3{ibid.\ #1 (19#3) #2}
\def\np#1#2#3{Nucl.\ Phys.\ B#1 (19#3) #2}
\def\pl#1#2#3{Phys.\ Lett.\ #1B (19#3) #2}
\def\pr#1#2#3{Phys.\ Rev.\ D #1 (19#3) #2}
\def\prep#1#2#3{Phys.\ Rep.\ #1 (19#3) #2}
\def\prl#1#2#3{Phys.\ Rev.\ Lett.\ #1 (19#3) #2}
\def\rmp#1#2#3{Rev.\ Mod.\ Phys.\ #1 (19#3) #2}
\def\sj#1#2#3{Sov.\ J.\ Nucl.\ Phys.\ #1 (19#3) #2}
\def\zp#1#2#3{Z.\ Phys.\ C#1 (19#3) #2}

\begin{enumerate}
\item \label{Catani:2000jh}
S.~Catani {\it et al.},
hep-ph/0005025, in the Proceedings of the CERN Workshop on 
{\it Standard
Model Physics (and more) at the LHC}, Eds. G. Altarelli and M.L.~Mangano, CERN 
2000-04, Geneva 2000.

\item \label{Catani:2000zg}
S.~Catani {\it et al.},
hep-ph/0005114, 
to be published in the Proceedings of the Les Houches Workshop on {\it
Physics at TeV Colliders}, Eds. P. Aurenche {\it et al.}

\item \label{AP}
G.\ Altarelli and G.\ Parisi, \np{126}{298}{77}.

\item \label{BCM}
A.\ Bassetto, M.\ Ciafaloni and G.\ Marchesini, \prep{100}{201}{83};
Yu.L.~Dokshitser, V.A.\ Khoze, A.H.\ Mueller and S.I. Troian,
{\it Basics of Perturbative QCD} (Editions Fronti\`eres, Gif-sur-Yvette, 1991)
and references therein.

\item \label{book}
R.K.\ Ellis, W.J.\ Stirling and B.R.\ Webber, {\it QCD and collider 
physics} (Cambridge University Press, Cambridge, 1996) and references therein.

\item \label{glover}
J.M.\ Campbell and E.W.N. Glover, \np{527}{264}{98}. 

\item \label{Catani:2000ss}
S.\ Catani and M.\ Grazzini,
Nucl.\ Phys.\  B570 (2000) 287.

\item \label{Bern:1995ix}
Z.\ Bern and G.\ Chalmers, \np{447}{465}{95}.

\item  \label{1loopepskos}
D.A.\ Kosower, \np{552}{319}{99}.
\item \label{Catani:1999nv}
S.~Catani and M.~Grazzini,
\pl{446}{143}{99}.

\item \label{DelDuca:2000ha}
V.~Del Duca, A.~Frizzo and F.~Maltoni,
Nucl.\ Phys.\  B568 (2000) 211.

\item \label{Bern:1998sc}
Z.~Bern, V.~Del Duca and C.R.~Schmidt,
Phys.\ Lett.\  B445 (1998) 168.

\item \label{Kosower:1999rx}
D.A.~Kosower and P.~Uwer,
Nucl.\ Phys.\  B563 (1999) 477.

\item \label{Bern:1999ry}
Z.~Bern, V.~Del Duca, W.B.~Kilgore and C.R.~Schmidt,
Phys.\ Rev.\  D60 (1999) 116001.

\item \label{bgdsoft}
F.A.\ Berends and W.T.\ Giele, \np{313}{595}{89}.

\item \label{sing2loop}
S.\ Catani, \pl{427}{161}{98}.

\item \label{1loopdec}
Z.~Bern and D.A.~Kosower,
Nucl.\ Phys.\  B362 (1991) 389;
Z.~Kunszt, A.~Signer and Z.~Tr\'ocs\'anyi,
Phys.\ Lett.\  B336 (1994) 529;
Z.~Bern, L.~Dixon and D.~A.~Kosower,
Nucl.\ Phys.\  B437 (1995) 259.

\item \label{tHV}
G. 't~Hooft and M.\ Veltman, \np{44}{189}{72}.

\item \label{bollini}
G.\ Bollini and J.J.\ Giambiagi, Nuovo\ Cimento\ 12B (1972) 20;
J.F.\ Ashmore, Nuovo\ Cimento\ Lett. 4 (1972) 289;
G.M.\ Cicuta and E.\ Montaldi, Nuovo\ Cimento\ Lett. 4 (1972) 329.

\item \label{Sie79}
W. Siegel, \pl{84}{193}{79} and 94B (1980) 37;
D.M.\ Capper, D.R.T.~Jones and P.\ van Nieuwenhuizen, \np{167}{479}{80};
L.V.\ Avdeev and A.A.~Vladi\-mirov, \np{219}{262}{83}.

\item \label{4dhs}
Z.\ Bern and D.A.\ Kosower, \np{379}{451}{92}.

\item \label{KST2to2}
Z.\ Kunszt, A.\ Signer and Z.\ Tr\'ocs\'anyi, \np{411}{397}{94}.

\item \label{uni}
S.\ Catani, M.H.\ Seymour and Z.\ Tr\'ocs\'anyi, \pr{55}{6819}{97}.

\item \label{CSdipole}
S.\ Catani and M.H.\ Seymour, \np{485}{291}{97}
(E ibid. B510 (1998) 503).

\item \label{YFS}
D.R.~Yennie, S.C.~Frautschi and H.~Suura,
Ann. Phys. 13 (1961) 379;
G.~Grammer and D.R.~Yennie,
Phys.\ Rev.\  D8 (1973) 4332.

\item \label{GG}
W.T. Giele and E.W.N. Glover, \pr{46}{1980}{92}.

\item \label{KS}
Z.\ Kunszt and D.E. Soper, \pr{46}{192}{92}.

\item \label{KST}
Z.\ Kunszt, A.\ Signer and Z. Tr\'ocs\'anyi, \np{420}{550}{94}.

\item \label{Catani:1985dp}
S.~Catani and M.~Ciafaloni,
Nucl.\ Phys.\   B249 (1985) 301.

\item \label{Catani:1986xt}
S.~Catani, M.~Ciafaloni and G.~Marchesini,
Nucl.\ Phys.\  B264 (1986) 588.

\item \label{CSdipolelet}
S.\ Catani and M.H.\ Seymour, \pl{378}{287}{96}.

\item \label{bassetto}
A.~Bassetto, G.~Nardelli and R.~Soldati,
{\it Yang-Mills theories in algebraic noncovariant gauges: 
Canonical quantization and renormalization},
(World Scientific, Singapore, 1991);
G.~Leibbrandt,
{\it Noncovariant gauges: Quantization of Yang-Mills and Chern-Simons theory 
in axial-type gauges} (World Scientific, Singapore, 1994).

\end{enumerate}

\end{document}